\documentclass[a4paper,11pt]{article}

\pdfoutput=1 
\usepackage{amsmath}
\usepackage{graphicx}

\usepackage[english]{babel}
\usepackage[utf8]{inputenc}
\usepackage{pdflscape}
\usepackage{physics}
\usepackage{enumerate}
\usepackage{amsbsy}
\usepackage{amsmath} 
\usepackage{graphics}
\usepackage{mathrsfs}
\usepackage{float}
\usepackage{wrapfig}
\usepackage{mathtools}
\usepackage{breqn}
\usepackage{amsfonts}
\usepackage{pstricks}
\usepackage{color}
\usepackage{setspace}
\usepackage{caption}
\usepackage{subcaption}
\usepackage{jcappub_ver} 

\renewcommand{\a}{\alpha}
\renewcommand{\b}{\beta}
\renewcommand{\c}{\gamma}

\newcommand{\pat}{\partial}

\title{Inflation and Reheating in $f(R, h)$ theory formulated  in  the Palatini formalism}
\author[a]{Nayan Das,}
\author[a]{Sukanta Panda}

\affiliation[a]{Department of Physics, Indian Institute of Science
	Education and Research, Bhopal 462066, India
}

\emailAdd{nayandas@iiserb.ac.in}
\emailAdd{sukanta@iiserb.ac.in}

\abstract{
	A new model for inflation using modified gravity in the Palatini formalism is constructed. 
Here non-minimal coupling of scalar field $h$ with the curvature $R$ as a general function $f(R,h)$ is considered. Explicit inflation models for some choices of $f(R,h)$ are developed. By writing an equivalent scalar-tensor action for this model and going over to Einstein frame,  slow roll parameters are constructed. There exists a large parameter space for different choices of $f(R,h)$ and potentials which satisfy values of $n_s$ and limits on $r$ compatible with Planck 2018 data. Further, we calculate reheating temperature and the number of e-folds at the end of reheating for different values of equation of state parameter for all the constructed models.
}

\begin{document}
	
	\maketitle
	
	\setcounter{tocdepth}{2}
	
	\setcounter{secnumdepth}{3}

\section{Introduction}

There are mainly two formulations in General Relativity (GR)  popularly known as Palatini and Metric formalisms. Palatini formalism or first order formalism treats space-time connections  as independent variable\cite{Sotiriou:2006hs,Sotiriou:2006qn,Sotiriou:2008rp,Borunda:2008kf,DeFelice:2010aj,Olmo:2011uz,Capozziello:2011et,Clifton:2011jh,  Nojiri:2017ncd}, whereas in Metric formalism, these connections are not independent but derived from the metric itself.  But in GR these two approaches produce same Einstein equation. Hence dynamics are equivalent in both formalisms. This is not true for modified gravity models and models where fields are nonminimally coupled to gravity. In these cases, both formalisms represent different physical situations \cite{Sotiriou:2006hs,Sotiriou:2006qn,Sotiriou:2008rp}.

Inflation \cite{PhysRevD.23.347, STAROBINSKY198099,LINDE1982389,LINDE1983177,Lyth:1998xn,Riotto:2002yw,Baumann:2009ds} was first developed in the early 1980s to solve problems of standard Big Bang theory like horizon problem, fine-tuning problem etc.  Quantum fluctuations also started during the period of inflation which led to the cosmic microwave background(CMB) anisotropy and provided the seed for  the formation of large scale structure of the universe. A more popular model for inflation is the Starobinsky model which is a pure gravity theory with an additional $R^2$ term in the Einstein-Hilbert action.  An equivalent scalar-tensor theory of Starobinsky model has an additional scalar degree of freedom apart from two tensor degrees of freedom of Einstein-Hilbert action\cite{Starobinsky:1980te,Mukhanov:1981xt,Starobinsky:1983zz}. In Einstein frame, this theory is equivalent to the usual scalar field model with a potential suitable for a valid inflation model which satisfies all the constraints from CMB data. 

In a broad sense, the Starobinsky model falls under a general framework of f(R) gravity. The above analysis works best in metric formulation of gravity. 
However, in the Palatini formalism, no additional propagating degrees of freedom appears in f(R) gravity theory \cite{Sotiriou:2008rp}. Because of this, no inflation is possible in this scenario. Hence a scalar field needs to be added to this action to develop an Inflation model in Palatini formulation of gravity. 
In this line, first work appeared in \cite{Bauer:2008zj} where nonminimal couplings of scalars are considered. Since then many works related to different inflationary potentials, preheating, reheating,  postinflationary phases, dark matter has been done with many variants of action including a $R^2$ term \cite{Koivisto:2005yc,Tamanini:2010uq,Bauer:2010jg,Enqvist:2011qm,Borowiec:2011wd,Stachowski:2016zio,Fu:2017iqg,Rasanen:2017ivk,Tenkanen:2017jih,Racioppi:2017spw,Markkanen:2017tun,Jarv:2017azx,Rasanen:2018ihz,Racioppi:2018zoy,Carrilho:2018ffi,Enckell:2018kkc,Bombacigno:2018tyw,Enckell:2018hmo,Antoniadis:2018ywb,Antoniadis:2018yfq,Rasanen:2018fom,Almeida:2018oid,Takahashi:2018brt,Kannike:2018zwn,Tenkanen:2019jiq,Shimada:2018lnm,Wu:2018idg,Kozak:2018vlp,Jinno:2018jei,Edery:2019txq,Rubio:2019ypq,Jinno:2019und,Giovannini:2019mgk,Tenkanen:2019xzn,Bostan:2019wsd,Tenkanen:2019wsd, Gialamas:2019nly, Shaposhnikov:2020fdv,Lloyd-Stubbs:2020pvx, Sa:2020qfd, Antoniadis:2020dfq,  Ghilencea:2020piz, Takahashi:2020car,Racioppi:2019jsp,Canko:2019mud,Myrzakulov:2015qaa}. For an introduction to Palatini inflation models refer to \cite{Tenkanen:2020dge} and references therein.  

In this work, we formulate a palatini inflation model in $f(R, h)$ theory. In metric formalism inflation model in $f(R, h)$ gravity has been considered earlier  in literature \cite{Rador:2007wq, delaCruz-Dombriz:2016bjj, Mathew:2017lvh}. Recently, it has been shown that the appearance of terms like $h^2R^2$ in the one loop effective action for massless, conformally coupled scalar field \cite{Glavan:2020ccz}. Our action is of the form similar to form obtained in \cite{Rador:2007wq}. But our approach here is to develop a model of inflation in palatini formalism. We show that such an action provide a favorable inflationary scenario satisfying Planck 2018 data \cite{Aghanim:2018eyx}.


Reheating \cite{Boyanovsky:1996sv,Bassett:2005xm,Allahverdi:2010xz,Amin:2014eta} is a phenomenon which acts as a transit between Inflation and the radiation dominated era of the universe. There are various models on how the inflaton field loses its energy. Using techniques similar to \cite{Dai:2014jja, Munoz:2014eqa, Cook:2015vqa, Drewes:2017fmn, Nautiyal:2018lyq}, the number of e-folds and temperature at the end of reheating ($T_{re}$ and $N_{re}$) can be written in terms of inflationary observables. Earlier, reheating in Palatini models of inflation has been discussed in these papers.
\cite{Gialamas:2019nly, Tenkanen:2019jiq, Lloyd-Stubbs:2020pvx}

This paper is organized as follows. In the next section, we introduce the action of our model.  Then we express it in terms of equivalent scalar-tensor action in Einstein frame. In section \ref{sec:slowroll}, slow roll parameters are defined. Values of $n_s$ and $r$ are also calculated numerically and the results are presented. Section \ref{sec:reheating} is dedicated to the calculations of reheating parameters. At last, in section \ref{sec:conclusion},
we summarize our results.

\section{The Model} \label{sec:mec}

We start with a general action of the form -

\begin{eqnarray} \label{action1}
S &=& \int\mathrm{d}^4 x \sqrt{-g} \left[  \frac{1}{2}f(R,h) - \frac{1}{2}g^{\mu \nu}  \pat_\mu h \pat_\nu h- V(h) \right] ,
\end{eqnarray}
where $g_{\a\b}$ is the metric, $g$ is its determinant, $f(R,h)=G(h)(R+\a R^2)$, $R$ is the Ricci scalar defined as $R=g^{\a\b} R^{\c}_{\ \, \a\c\b}(\Gamma, \pat\Gamma)$ in Palatini formalism and $\a$ is a constant. Here we have chosen Planck mass to be unity. An equivalent action in terms of an auxiliary field $\phi$ can be written as -
\begin{eqnarray}\label{action2}
S &=& \int\mathrm{d}^4 x \sqrt{-g} \left[  \frac{1}{2} f(\phi,h) +  \frac{1}{2}\ f'(\phi,h) (R - \phi) \ - \frac{1}{2} g^{\mu \nu} \pat_\mu h \pat_\nu h - V(h) \right] \  ,
\end{eqnarray} 
where $f'(\phi,h)=\frac{\partial f(\phi,h)}{\partial \phi}$ . Varying $S$ with respect to $\phi$ in equation \eqref{action2}, we get $\phi = R$ if $\pdv[2]{f(\phi,h)}{ \phi}\ne 0$ and using this result in equation \eqref{action2}, we recover equation \eqref{action1}. Rearranging equation \eqref{action2}, the action becomes -
\begin{eqnarray}\label{action3}
S &=& \int\mathrm{d}^4 x \sqrt{-g} \left[  \frac{1}{2} f'(\phi,h) R - W(\phi,h)- \frac{1}{2} g^{\mu \nu} \pat_\mu h \pat_\nu h - V(h) \right] \  ,
\end{eqnarray}
where $W(\phi,h)= \frac{1}{2}\phi f'(\phi,h) - \frac{1}{2} f(\phi,h)$. By making a conformal transformation -
\begin{eqnarray}
g_{\mu \nu} \rightarrow f'(\phi,h)  g_{\mu \nu} \ ,
\end{eqnarray}
action is obtained in Einstein frame as -
\begin{eqnarray}\label{action4}
S &=& \int\mathrm{d}^4 x \sqrt{-g} \left[  \frac{1}{2} R -\frac{1}{2}\frac{ \pat_\mu h \pat^\mu h}{f'(\phi,h)}- \frac{W(\phi,h)+V(h)}{f'(\phi,h)^2} \right] \  .
\end{eqnarray}
Let us define a new potential $\hat V(\phi,h)$ as -
\begin{eqnarray}\label{V}
\hat V(\phi,h)\equiv \frac{W(\phi,h)+V(h)}{f'(\phi,h)^2} .
\end{eqnarray}
Now for our choice of $f(R,h) = G(h)(R+\a R^2)$\cite{Rador:2007wq}, the new potential can be written as -
\begin{eqnarray}\label{V1}
\hat V(\phi,h)= \frac{1}{f'(\phi,h)^2} \left[\frac{1}{8 \a G(h)}[f'(\phi,h)-G(h)]^2 +V(h)\right] .
\end{eqnarray}
Varying equation \eqref{action4} with respect to $\phi$, we get the constraint equation -
\begin{eqnarray}\label{V2}
f'(\phi,h)= \frac{8\a V(h) + G(h)}{1- 2\a \pat_\mu h \pat^\mu h} .
\end{eqnarray}
Inserting equation \eqref{V2} to equation \eqref{V1}, we can eliminate $\phi$. Then again inserting equation \eqref{V1} to equation \eqref{action4} and rearranging and simplifying, we get -
\begin{dmath} \label{action5}
	S = \int\mathrm{d}^4 x \sqrt{-g} \left [ \frac{1}{2} R - \frac{1}{2}\frac{1}{(8 \a V + G )} \pat^\mu h \pat_\mu h + \frac{\a}{2} \frac{1}{(8\a V+G)} ( \pat^\mu h \pat_\mu h )^2 - \frac{V}{G(8\a V+G)}\right ] \ .
\end{dmath} 
Now the last term can be defined as effective potential in the Einstein frame -
\begin{eqnarray} \label{U}
U\equiv  \frac{V}{G(8\a V+G)} \ .
\end{eqnarray} 
In order to bring kinetic part of scalar field into that of canonical form, we introduce a new field $\chi$ as -
\begin{eqnarray} \label{C}
\dv{h}{\chi} = \pm \sqrt{(8 \a V + G )} .
\end{eqnarray}
In terms of $\chi$, equation \eqref{action5} becomes -
\begin{eqnarray} \label{action6}
S = \int\mathrm{d}^4 x \sqrt{-g} \left [ \frac{1}{2} R - \frac{1}{2} \pat^\mu \chi \pat_\mu \chi + \frac{\a}{2}{(8\a V+G)} ( \pat^\mu \chi \pat_\mu \chi )^2 -U\right ] \ .
\end{eqnarray} 

Now we are ready to build up our inflation model, with this action.  Next, we define the slow-roll parameters to estimate the observables which will decide the fate of our model.  Here we have neglected contribution due to the third term of the action to the inflation phase. This is a valid assumption so far as the slow-roll inflation is concerned (see \cite{Tenkanen:2020dge} and references therein). In this case then, it is a model of inflation driven by a scalar field with potential $U$ \footnote{Background evolution of $\chi$ field for a flat FRW metric with scale factor $a$ and hubble parameter $H=\frac{\dot{a}}{a}$ is-
	\begin{eqnarray} \label{perturbedeom}
	a^2 \ddot{\chi} + 2 a^3 H \dot{\chi} + 12 a H f(\chi) \left(\dot{\chi}\right)^2 + 3 \frac{df(\chi)}{d\chi}\left(\dot{\chi}\right)^4
	+ a^4 U'=0
	\end{eqnarray}
	where $f(\chi)=\frac{\a}{2}(8 \a V + G)$ and $\dot{}$ and $'$ represent derivatives with respect to conformal time and $\chi$ respectively. Under slow-roll approximation, terms proportional to $\dot{\chi}^2$ and   $\dot{\chi}^4$ can be neglected and equation \eqref{perturbedeom}
reduces to usual FRW equations for minimally coupled field.
	}.

\section{Slow-roll parameters and Results} \label{sec:slowroll}

Before analyzing the dynamics of our model, we define slow-roll parameters (mainly $\epsilon$ and $\eta$) which are helpful to decide whether a model can describe inflation or not. They can be written as -
\begin{eqnarray}\label{ep}
\epsilon &=& \frac{1}{2} \left (\frac{\frac{dU}{d\chi}}{U}\right)^2 , \nonumber \\
\eta &=& \frac{\frac{d^2 U}{d \chi^2}}{U} .
\end{eqnarray}
 The slow roll parameters $\epsilon$ and $\eta$ must be $\ll 1$ during inflation phase of expansion. For the potential in equation \eqref{U}, slow roll parameters become -

\begin{eqnarray}\label{e}
\epsilon &=& \frac{G\bar{\epsilon}}{(8\a V+G)} \left [1-\left(\frac{8 \a \frac{V}{G^2}G'}{\frac{V'}{V}-\frac{2G'}{G}}\right)\right]^2  ,\nonumber \\
\eta &=& \bar{\eta}\frac{G}{8\a V+G}- \left [(-256 \a^2 V^3 {G'}^2+8 \a G V^2(-11 {G'}^2+8 \a G' V' +16 \a V G'')\right. \nonumber \\ &&   \left.+2 G^2 V(8 \a G' V' +24 \a V G'') + G^3(24\a {V'}^2-16 \a V V''))/(2 G^2 V(8 \a V+G))\right] ,
\end{eqnarray}
where $\bar{\epsilon}$ and $\bar{\eta}$ are slow-roll parameters for $\a = 0$ and $'$ represents derivative with respect to $h$. Number of e-folds from the time when a mode $k$ crosses the horizon to the end of inflation, $N_k$ can be written as (for unit Planck mass) -
\begin{eqnarray} \label{N}
N_k = - \int_{h_k}^{h_{end}} \mathrm{d}h \frac{1}{\pm \sqrt{2 \epsilon (h) (8 \a V+G)}} .
\end{eqnarray}
Inflation ends when $\epsilon \simeq 1$ and this, in turn, determines the value of $h_{end}$. Putting the value of $h_{end}$  and $N_k$ (taken in between 60 and 70) in \eqref{N} , value of $h_k$ can be obtained. Putting this value to equations \eqref{ep} and \eqref{e}, value of $\epsilon$ and $\eta$ can be obtained respectively. Scalar spectral index and tensor-to-scalar ratio can then be calculated from the following equations -
\begin{eqnarray}\label{S}
r= 16\epsilon , \nonumber \\
n_s=1+ 2\eta-6\epsilon .
\end{eqnarray}
Now we estimate $r$ and $n_s$  for various cases. Here we restrict our analysis only for quadratic and quartic potentials. We choose two values of $G=\gamma h^2$ and $G=1+\gamma h^2$ and $\gamma$ is a dimensionless constant.  

\subsection*{Case 1 : $V=\b h^2$ and $G=\c h^2$} 
For this case -
\begin{eqnarray}
\epsilon &=& 2 ( 8 \a \b +\c) , \nonumber \\
\eta &=& 4 (8 \a \b + \c)  .
\end{eqnarray}
As neither $\epsilon$ nor $\eta$ depend on $h$, so it will lead to constant  $r$ and $n_s$ for a particular choice
of $\a$, $\b$ and $\c$.  In this case, then it would be difficult to end the inflation and to decide the value of $h_f$. Therefore 
we will not discuss this case further here.

\subsection*{Case 2 : $V=\b h^4$ and $G=\c h^2$} 
\begin{figure}[H]
	\centering
	\begin{minipage}{.5\textwidth}
		\centering
		\includegraphics[width=7cm]{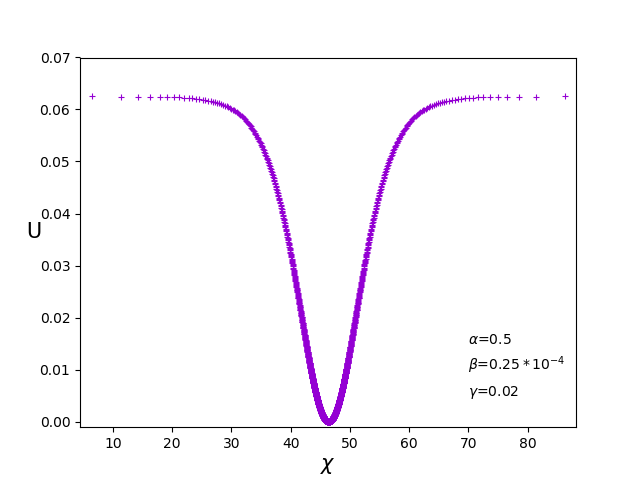}
		\caption{Potential as a function of canonical scalar field}
		\label{fig:c2cvsu}
	\end{minipage}%
	\begin{minipage}{.5\textwidth}
		\centering
		\includegraphics[width=7cm]{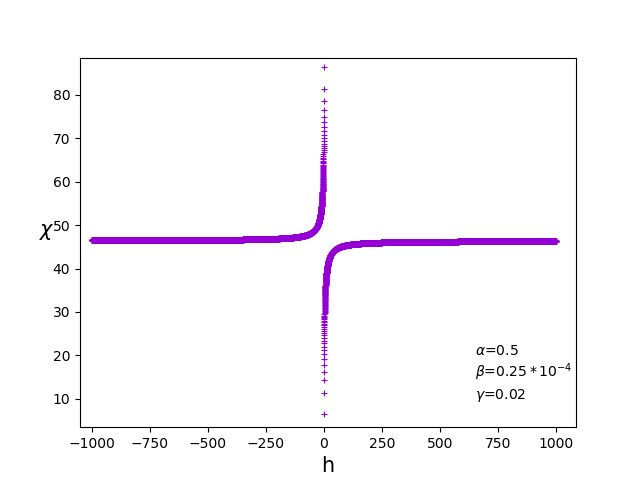}
		\caption{Plot of $\chi(h)$}
		\label{fig:c2hvsc}
	\end{minipage}
	\begin{minipage}{.5\textwidth}
		\centering
		\includegraphics[width=8cm]{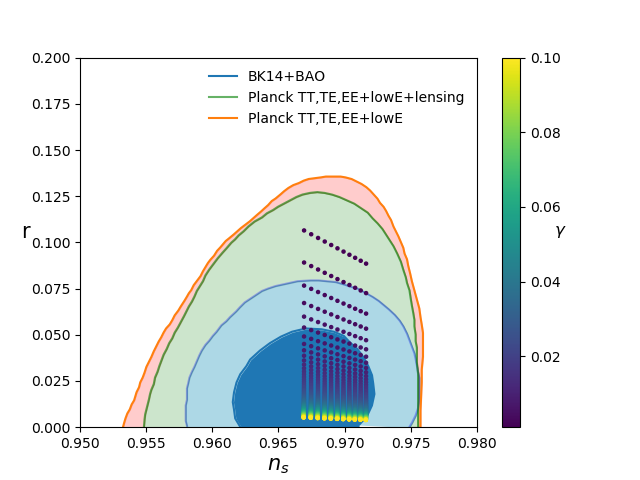}
		\caption{Plot of $r$ vs $n_s$ for changing $\c$}
		\label{fig:plot1}
	\end{minipage}
\end{figure}

In this case, $\epsilon$ and $\eta$ depend only on h and $\c$ for a particular value of $N$. The parameters $\a$ and $\b$ fixes the value of the field $\chi$ or $h.$  In fig. \ref{fig:c2cvsu}, potential $U$ vs $\chi$ is plotted. In fig. \ref{fig:c2hvsc} the relation between $h$ vs $\chi$ is plotted. The value of $\chi$ changes discontinuously as $h$ changes from negative to positive value. Here positive sign is taken in equation \eqref{C}. Instead if negative sign is taken, all values of $\chi$'s will be negative. However, discontinuity of $\chi$ is there for both positive and negative sign. For fig. \ref{fig:c2cvsu} and fig. \ref{fig:c2hvsc}, the value of parameters $\a = 0.5$, $\b = \frac{1}{4}\times 10^{-4} $ and $\c = 0.02$ is chosen. For this particular set, the numerically estimated  value of $n_s $ and $r$ are $0.967 $ and $0.023$ respectively(we take $N=60$ here). Also the result for spectral index and tensor to scalar ratio is shown in fig. \ref{fig:plot1}. Here $\c$ is changing and $\a=5$, $\b=\frac{1}{4}\times 10^{-4}$ and negative sign is considered in equation \eqref{C}. It clearly shows that there can be large parameter space available which can satisfy Planck constraints.

\subsection*{Case 3 : $V=\b h^4$ and $G=1+\c h^2$} 
\begin{figure}[H]
	\begin{minipage}{.5\textwidth}
		\centering
		\includegraphics[width=7cm]{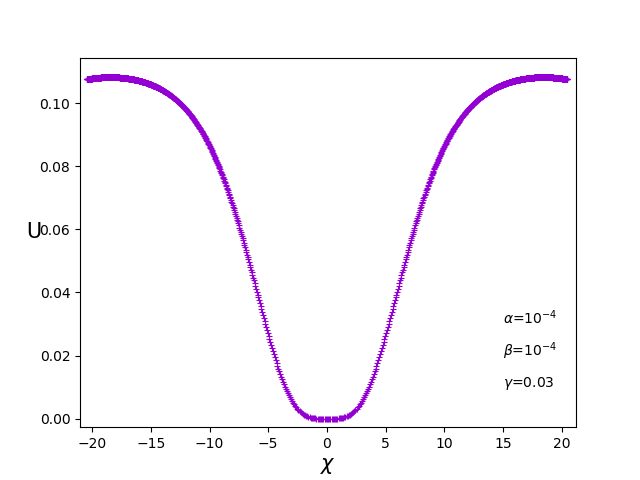}
		\caption{Potential as a function of canonical scalar field}
		\label{fig:c3cvsu}
	\end{minipage}%
	\begin{minipage}{.5\textwidth}
		\centering
		\includegraphics[width=7cm]{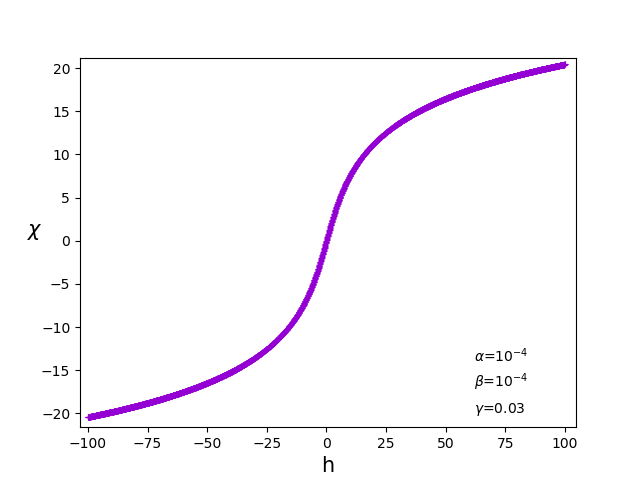}
		\caption{Plot of $\chi(h)$}
		\label{fig:c3hvsc}
	\end{minipage}
	\begin{minipage}{.5\textwidth}
		\centering
		\includegraphics[width=8cm]{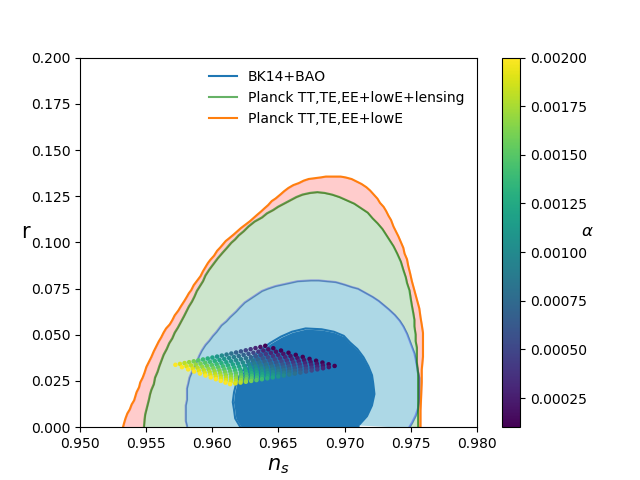}
		\caption{Plot of $r$ vs $n_s$ for changing $\a$}
		\label{fig:plot2}
	\end{minipage}%
	\begin{minipage}{.5\textwidth}
		\centering
		\includegraphics[width=8cm]{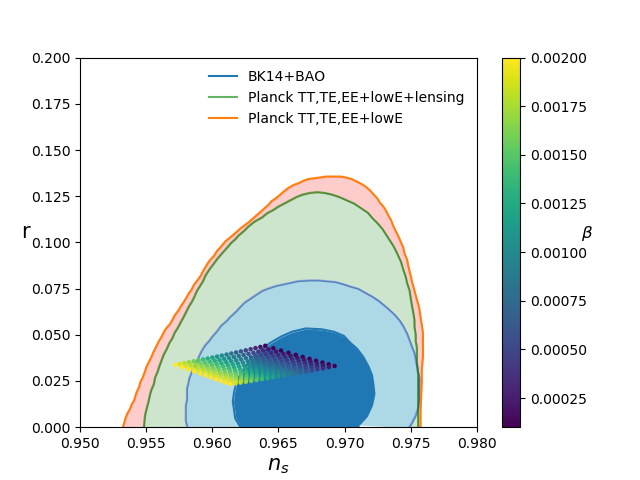}
		\caption{Plot of $r$ vs $n_s$ for changing $\b$}
		\label{fig:plot2between3f}
	\end{minipage}
\end{figure}

\begin{figure}[H]
	\begin{minipage}{.5\textwidth}
		\centering
		\includegraphics[width=8cm]{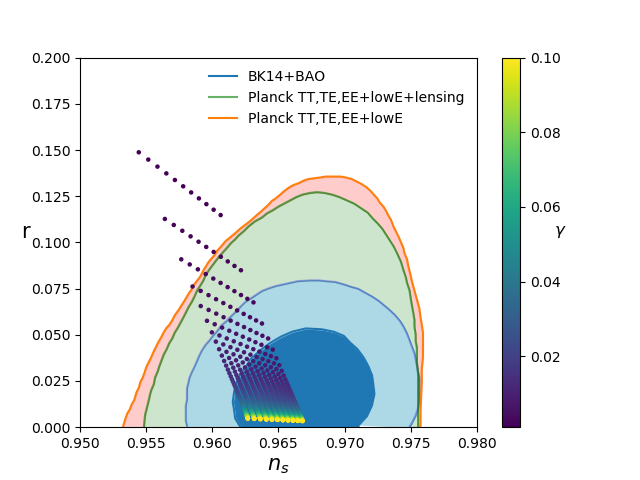}
		\caption{Plot of $r$ vs $n_s$ for changing $\c$}
		\label{fig:plot3}
	\end{minipage}
\end{figure}

For case 3 and case 4, $G(h)$ is taken as $G=1+\c h^2$. This makes the Lagrangian in \eqref{action1} equivalent to the form $\frac{1}{2}(R+\a R^2) + \frac{1}{2} \c h^2(R+\a R)$ plus usual scalar field term. Similar to case 2, the graphs of $U$ vs $\chi$ and $h$ vs $\chi$ are shown in fig \ref{fig:c3cvsu} and fig. \ref{fig:c3hvsc} respectively. Unlike to the previous case, here $\chi(h)$ is continuous for all range of $h$.
For fig \ref{fig:c3cvsu} and fig \ref{fig:c3hvsc}, the value of parameters are $\a = 10^{-4}$, $\b = 10^{-4}$ and $\c = 0.03.$ For this particular value of parameters, we obtain $n_s = 0.966$ and $r = 0.017$. The potential in equation \eqref{U} takes the form -
\begin{eqnarray} \label{c3U}
U=\frac{\b h^4}{(1 + \c h^2)(8 \a \b h^4 + 1 + \c h^2)}.
\end{eqnarray}
Dividing both numerator and denominator by $h^6$, we get -
\begin{eqnarray}
U=\frac{\b \frac{1}{h^2}}{(\frac{1}{ h^2} + \c)(8 \a \b  + \frac{1}{ h^4} + \c \frac{1}{h^2})} .
\end{eqnarray}
We can clearly observe that $U \to 0$ as $h \to \infty $ . Hence this potential will not give rise to plateau unlike the results for Lagrangian where $h$ is coupled only with $R$ \cite{Enckell:2018hmo,Antoniadis:2018ywb,Antoniadis:2018yfq,Tenkanen:2020cvw, Tenkanen:2019jiq,Jinno:2018jei}. The results for $n_s$ and $r$  are shown in fig. \ref{fig:plot2}, \ref{fig:plot2between3f} and \ref{fig:plot3}. In fig. \ref{fig:plot2}, $\b=10^{-4}$, $\c=10^{-2}$ and $\a$ is changing. In fig. \ref{fig:plot2between3f}, $\a=10^{-4}$, $\c=10^{-2}$ and $\b$ is changing. And  in fig. \ref{fig:plot3}, $\a=10^{-3}$, $\b=10^{-4}$ and $\c$ is changing. Here we consider positive sign in equation \eqref{C}. Note that both \ref{fig:plot2} and \ref{fig:plot2between3f} look same. This happens because in the expression of $r$ and $n_s$, $\a$ and $\b$ always appear in pairs i.e. in the form of $(\a\b)$ or ${(\a\b)}^2$. Similar to case 2,  here also a large parameter space available which satisfies Planck constraints on $r$ and $n_s.$  It is observed from \ref{fig:plot3} that we can tune the parameters of our model to obtain a very small value for $r.$  
\subsection*{Case 4 : $V=\b h^2$ and $G=1+\c h^2$} 
\begin{figure}[H]
	\begin{minipage}{.5\textwidth}
		\centering
		\includegraphics[width=7cm]{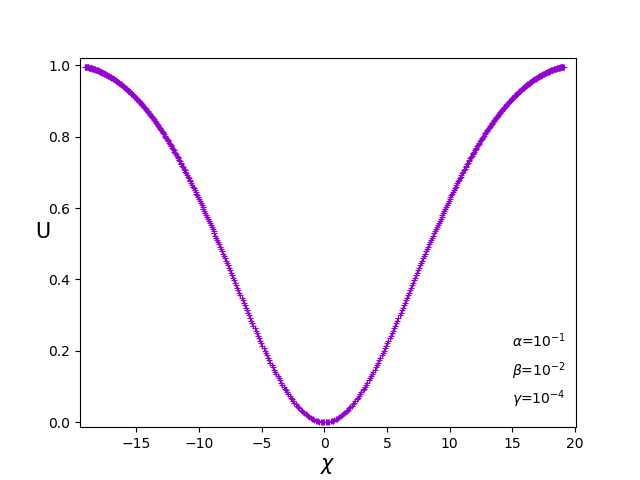}
		\caption{Potential as a function of canonical scalar field}
		\label{fig:c4cvsu}
	\end{minipage}%
	\begin{minipage}{.5\textwidth}
		\centering
		\includegraphics[width=7cm]{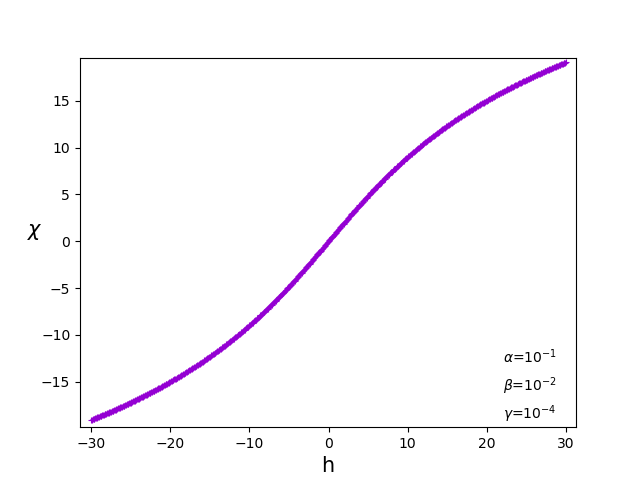}
		\caption{Plot of $\chi(h)$}
		\label{fig:c4hvsc}
	\end{minipage}
\end{figure}

\begin{figure}[H]
	\begin{minipage}{.5\textwidth}
		\centering
		\includegraphics[width=8cm]{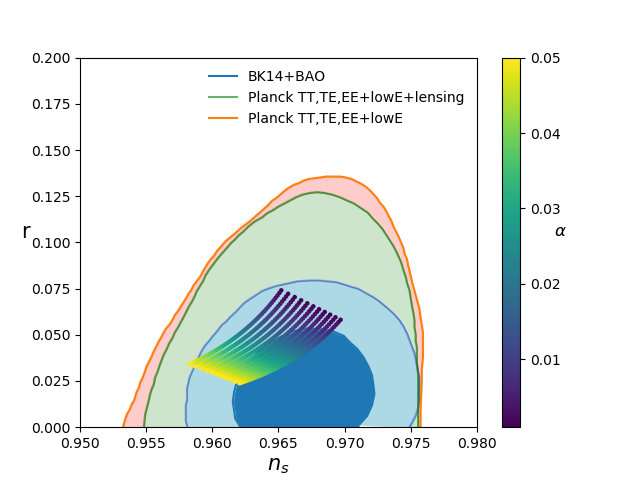}
		\caption{Plot of $r$ vs $n_s$ for changing $\a$}
		\label{fig:plot4}
	\end{minipage}%
	\begin{minipage}{.5\textwidth}
		\centering
		\includegraphics[width=8cm]{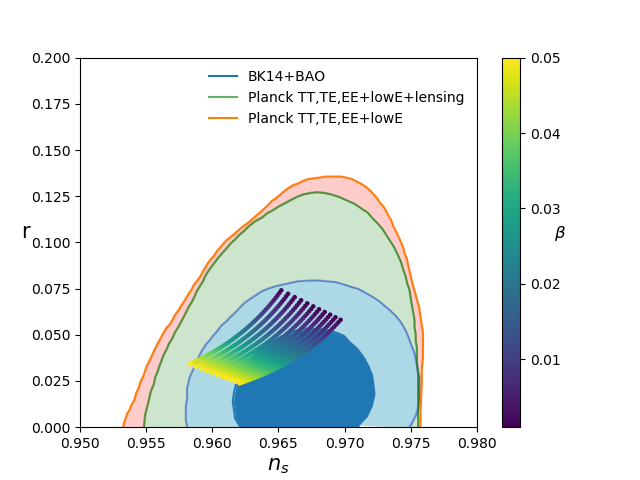}
		\caption{Plot of $r$ vs $n_s$ for changing $\b$}
		\label{fig:plot5}
	\end{minipage}
\end{figure}

\begin{figure}[H]
	\begin{minipage}{.5\textwidth}
		\centering
		\includegraphics[width=8cm]{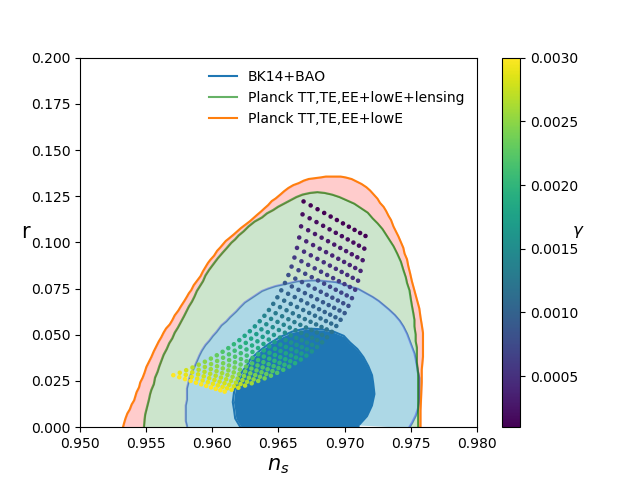}
		\caption{Plot of $r$ vs $n_s$ for changing $\c$}
		\label{fig:plot6}
	\end{minipage}
	
\end{figure}

Analogous to case 3, potential graph and $h$ vs $\chi$ are shown in fig \ref{fig:c4cvsu} and fig \ref{fig:c4hvsc} for $\a = 0.1$, $\b = 0.01$ and $\c = 10^{-4}$ which give $n_s=0.965$ and $r=0.04$. Here, potential in equation \eqref{U} becomes -
\begin{eqnarray} \label{Ucae4}
U = \frac{\b h^2}{(1 + \c h^2)(8 \a \b h^2 + 1 + \c h^2)} .
\end{eqnarray}
As $h \to \infty$, $U \to 0$. So again no plateau in the potential. The $n_s$ and $r$ graphs are plotted in fig. \ref{fig:plot4}, \ref{fig:plot5} and \ref{fig:plot6}. In fig. \ref{fig:plot4} $\b = 0.01$, $\c = 0.001$ and $\a$ is changing. In fig. \ref{fig:plot5} $\a=10^{-2}$, $\c=10^{-3}$ and $\b$ is varying. In fig. \ref{fig:plot6} $\a = 10^{-3}$, $\b = 0.01$ and $\c$ is varying.  Here we consider positive sign in equation \eqref{C}. In this case as well, we find a large parameter space available which is compatible with Planck 2018 data.


\section{Reheating}  \label{sec:reheating}
In this section our aim is to find the reheating temperature, $T_{re}$ and number of e-folds at the end of reheating, $N_{re}$ for our models. Equation of state during reheating is parameterised by a function $\omega(t)$, which is obtained from - 
\begin{eqnarray}
P=\rho  \omega(t) ,
\end{eqnarray}
where $P$ and $\rho$ represent pressure and density of a particular component. For radiation and matter dominated universe, the value of $\omega $ is $1/3$ and $0$ respectively. Here we consider $\omega_{re}$ ranging from $-1/3$ to $1$ during reheating period \cite{Cook:2015vqa}. We also assume that $\omega_{re}$ is constant throughout the reheating period.
Mathematically, the number of e-folds at the end of reheating is defined as -
\begin{eqnarray} \label{nre}
N_{re}= \ln(\frac{a_{re}}{a_{end}}) ,
\end{eqnarray}
where $a_{re}$ denotes scale factor at the end of reheating and $a_{end}$ denotes scale factor at the end of inflation. Using continuity equation along with $k=a_k H_k$ and assuming conservation of energy, one can derive the following equations \cite{Cook:2015vqa}-
\begin{eqnarray} \label{tre}
T_{re}=\left(\frac{43}{11 g_{re}}\right)^{\frac{1}{3}}\left(\frac{a_0 T_0}{k}\right)H_k e^{-N_k} e^{-N_{re}} ,
\end{eqnarray}
\begin{align} \label{nre1}
N_{re}= \frac{4}{3(1+\omega_{re})}\left[\frac{1}{4}\ln(\frac{45}{\pi^2 g_{re}})+\ln(\frac{V_{end}^{\frac{1}{4}}}{H_k})+\frac{1}{3}\ln(\frac{11 g_{re}}{43})+\ln(\frac{k}{a_0 T_0})+N_k+N_{re} \right] .
\end{align}
Here $k$ denotes pivot scale of horizon crossing, $V_{end}$ denotes the potential at the end of inflation, and $g_{re}$ denotes the number of relativistic species at the end of reheating. Note that in equation \eqref{tre}, as $N_{re}$ increases, $T_{re}$ decreases. Hence more instant reheating implies more temperature. For $\omega_{re}=\frac{1}{3}$, we can not extract any information of $N_{re}$ from equation \eqref{nre1} as both RHS and LHS get canceled out. This happens because when $\omega_{re}=\frac{1}{3}$, the boundary between reheating period and the radiation dominated era is indistinguishable. For $\omega_{re}=\frac{1}{3}$, equation \eqref{nre1} becomes -
\begin{align} \label{nre2}
0= \frac{1}{4}\ln(\frac{45}{\pi^2 g_{re}})+\ln(\frac{V_{end}^{\frac{1}{4}}}{H_k})+\frac{1}{3}\ln(\frac{11 g_{re}}{43})+\ln(\frac{k}{a_0 T_0})+N_k .
\end{align}
For $\omega_{re}\neq \frac{1}{3}$, equation \eqref{nre1} writes as-
\begin{align} \label{nre3}
N_{re}= \frac{4}{(1-3\omega_{re})}\left[-\frac{1}{4}\ln(\frac{45}{\pi^2 g_{re}})-\ln(\frac{V_{end}^{\frac{1}{4}}}{H_k})-\frac{1}{3}\ln(\frac{11 g_{re}}{43})-\ln(\frac{k}{a_0 T_0})-N_k \right].
\end{align}
Assuming $g_{re}\approx 100$ and $k=0.05$ Mpc$^{-1}$ \cite{Cook:2015vqa}, equations \eqref{nre3} and \eqref{tre} become -

\begin{eqnarray} \label{nre4}
N_{re}= \frac{4}{1-\omega_{re}}\left[61.6-\ln(\frac{V_{end}^{\frac{1}{4}}}{H_k})-N_k \right] ,
\end{eqnarray}
\begin{eqnarray} \label{tre1}
T_{re}=\left[\left(\frac{43}{11 g_{re}}\right)^{\frac{1}{3}}\frac{a_0 T_0}{k} H_k e^{-N_k}\left[\frac{45 V_{end}}{\pi^2 g_{re}}\right]^{\frac{-1}{3(1+\omega_{re})}}\right]^{\frac{3(1+\omega_{re})}{3 \omega_{re}-1}} .
\end{eqnarray}

Until now, we have done the model independent calculations. Now we are ready to do the model dependent calculations for our cases except case $1$.

\subsection*{Case 2 : $V=\b h^4$ and $G=\c h^2$} 
\begin{figure}[h]
	\centering
	
	\includegraphics[width=7cm]{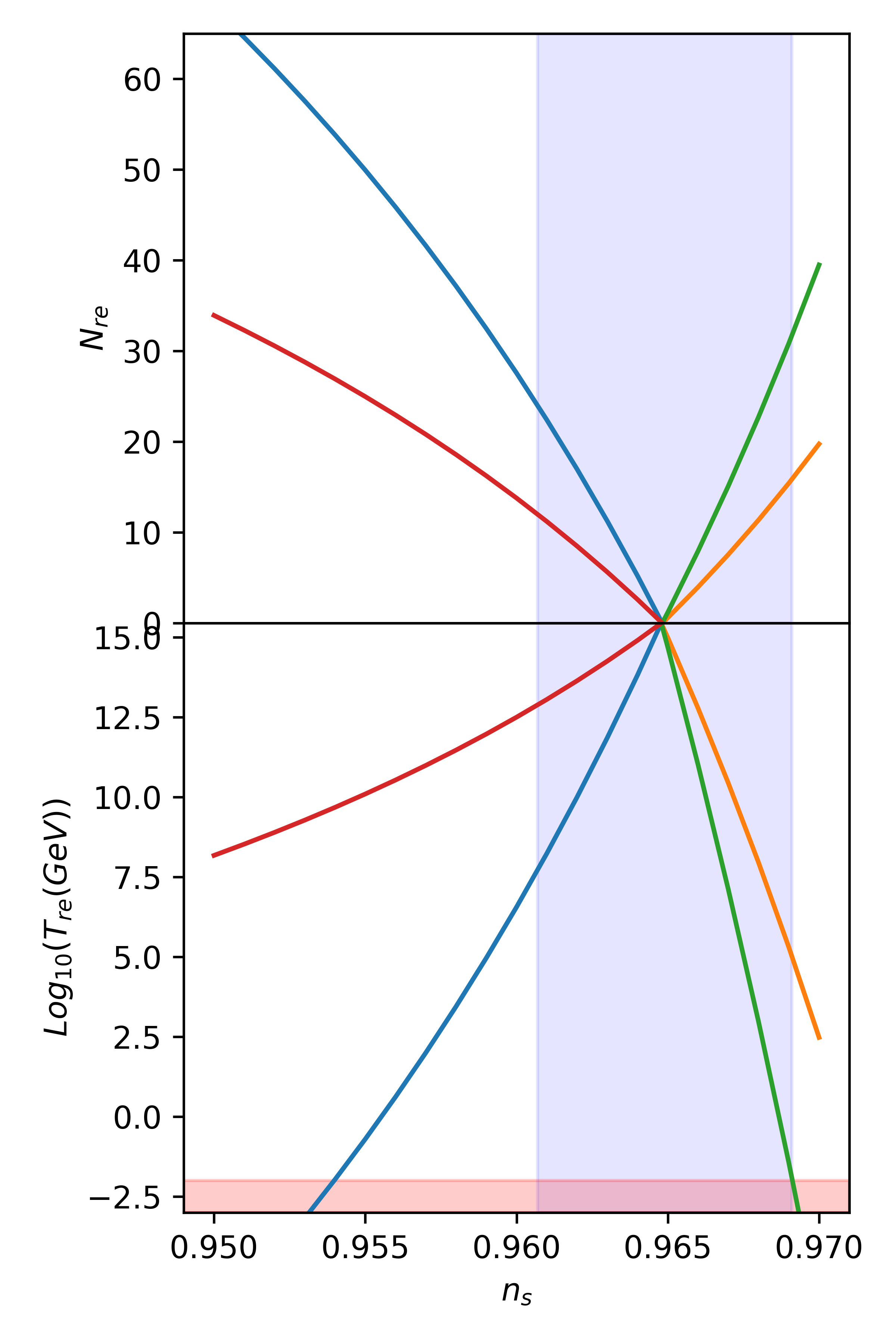}
	
	\captionsetup{justification=centering,margin=2cm}
	
	\caption{ Plots of $N_{re}$ and $T_{re}$ with respect to $n_s$. Here red, blue, green and orange colour lines correspond to $\omega_{re}=-\frac{1}{3}, 0, \frac{2}{3}$ and $1$ respectively.  }
	\label{fig:reheatingcase2}
	
\end{figure}

For this case, equation \eqref{N} provides (considering $-$ sign before $\sqrt{2 \epsilon (h) (8 \a V+G)}$ term) -
\begin{eqnarray}
N_k=\frac{1}{32 \a \b h_k^2}-\frac{1}{32 \a \b h_{end}^2} .
\end{eqnarray}
Assuming $h_{k}<<h_{end}$, we get $N_k\approx\frac{1}{128 \a \b h_{k}^2}$. Also $\epsilon$ and $\eta$ for this case is written as-
\begin{eqnarray}
\epsilon=\frac{128 \a^2 \b^2 h^4}{8 \a \b h^2 + \c} ,
\end{eqnarray}
\begin{eqnarray}
\eta=\frac{128 \a^2 \b^2 h^4 - 32 \a \b h^2}{8 \a \b h^2 + \c} .
\end{eqnarray}
Using the above relations, we obtain-
\begin{eqnarray} \label{nk}
N_k=\frac{2}{1-n_s},
\end{eqnarray}
\begin{eqnarray}
\epsilon_k=\frac{(1-n_s)^2}{4(1-n_s)+32 \c},
\end{eqnarray}
\begin{eqnarray}
V_{end}=3 M_p^2 H_k^2 \frac{\frac{1-n_s}{8}+\c}{8 \a \b h_{end}^2 + \c}.
\end{eqnarray}
 Also $H_k$ can be written as -
\begin{eqnarray}
H_k=\pi M_p \sqrt{8 A_s \epsilon_k} .
\end{eqnarray}
Equations \eqref{nre4} and \eqref{tre1} can be expressed in terms of inflationary observable parameter $n_s$ using the above relations. Figure \ref{fig:reheatingcase2} is obtained by varying $n_s$ from $0.95$ to $0.97$ for $\a= 0.5$, $\b= 0.25\times10^{-4}$ and $\c=0.01$. Here we use $A_s=2.196\times10^{-9}$. The shaded blue region corresponds to Planck's $n_s$ value and the shaded red region is obtained from the fact that Big Bang nucleosynthesis temperature should not be less than $10^{-2}  GeV$. All lines meet when $N_{re}\approx0$  which means that instantaneous reheating occurs irrespective of the choices of $\omega_{re}$.  Also, the temperature is maximum for instantaneous reheating as expected. For $\omega_{re}=\frac{1}{3}$, as any value of $N_{re}$ would satisfy equation \eqref{nre2}, hence it will give rise to a vertical line passing through instantaneous reheating point. The graph shows good agreement with Planck's data for all values of $\omega_{re}$.

\subsection*{Case 3 : $V=\b h^4$ and $G=1+\c h^2$} 

For case 3, $N_k$ can not be expressed in terms of $n_s$ in a simplified form. Hence we need to take some approximation. Note that in fig. \ref{fig:c3hvsc}, we have taken $\a=10^{-4}$, $\b=10^{-4}$ and $\c=0.03$ which are favourable by Planck data. Also value  $h_k$ is of the  $O(10)$. Hence, in the equation \eqref{c3U}, we can safely neglect $ 8 \a\b h^4$ term in comparison to $1$ and $\c h^2$. With this approximation, $\epsilon$, $\eta$, $n_s$ and $N_k$ become -

\begin{eqnarray}
	\epsilon\simeq \frac{8}{h^2(1 + \c h^2)},
\end{eqnarray}

\begin{eqnarray}
\eta \simeq \frac{12}{h^2} - \frac{20 \c}{(1 + \c h^2)},
\end{eqnarray}

\begin{eqnarray}
 N_k \simeq \frac{h^2_k}{8}.
\end{eqnarray}

Using the above expressions, we obtain the following equations -

\begin{eqnarray} \label{rcase3nk}
	N_k \simeq \frac{-1 + n_s + 16 \c + \sqrt{1 - 2 n_s + n^2_s + 64 \c - 64 \c n_s + 256 \c^2}}{16 (\c - \c n_s)} ,
\end{eqnarray}

\begin{eqnarray}
	n_s \simeq 1 - \frac{2 + \sqrt{1 + \frac{32 \c}{\epsilon}}}{\epsilon}  ,
\end{eqnarray}

\begin{eqnarray}
	V_{end} \simeq 3 H^2_k \frac{(1 + 8 \c N_k)(512 \a \b N^2_k + 1 + 8 \c N_k) h^2_{end}}{(1 + \c h^2_{end})(8 \a \b h^4_{end} + 1 + \c h^2_{end})  8 N_K} .
\end{eqnarray}

\begin{figure}[H]

	\begin{minipage}{.5\textwidth}

		\includegraphics[width=7cm]{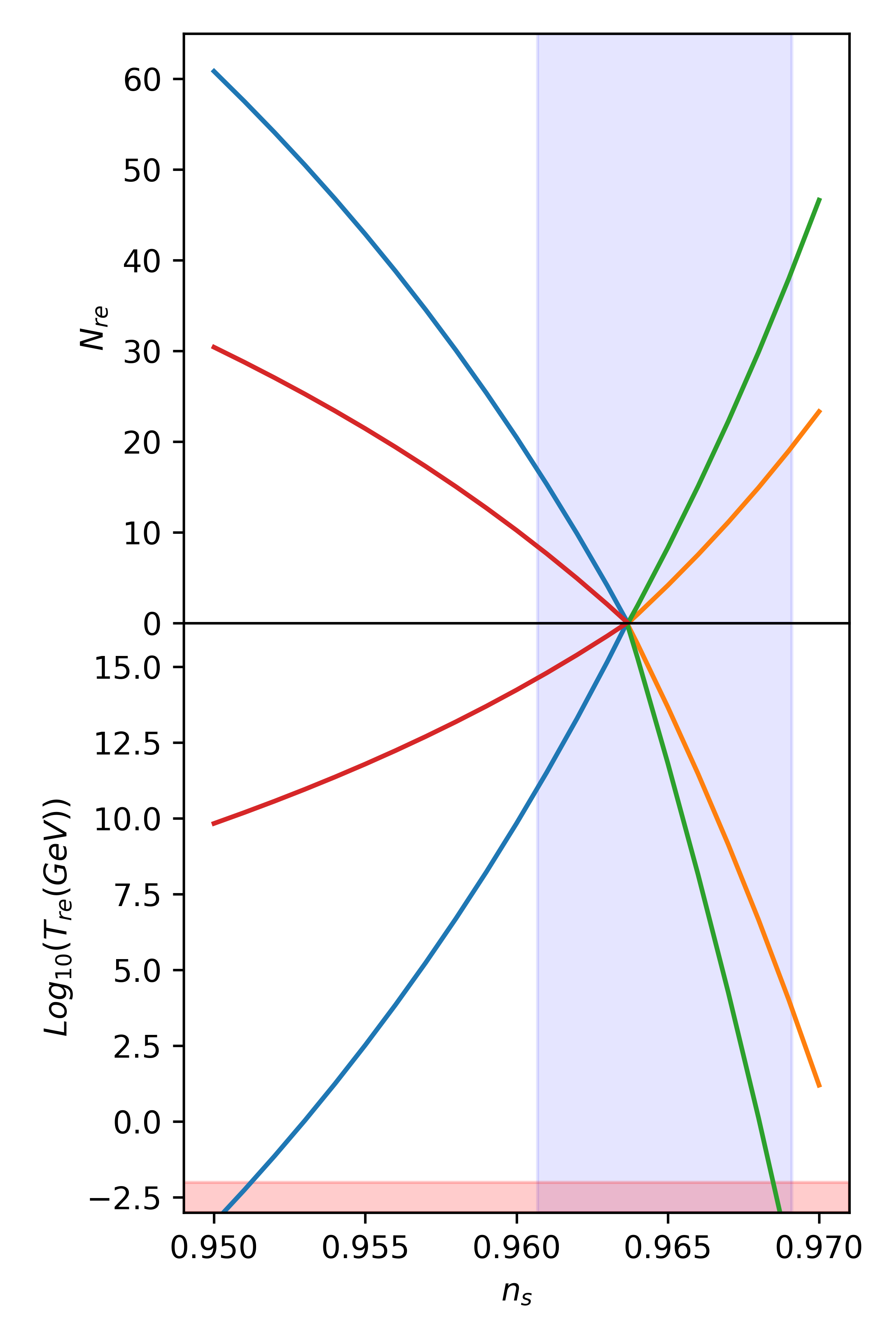}
		
		\captionsetup[figure]{position=bottom,justification=centering,width=.85\textwidth,labelfont=bf,font=small}
		
		\caption{ Plots of $N_{re}$ and $T_{re}$ with respect \\
			to $n_s$. Here red, blue, green and orange colour 
		\\
		 lines correspond to $\omega_{re}=-\frac{1}{3}, 0, \frac{2}{3}$ and $1$ \\
		 respectively.  }
		\label{fig:reheatingcase3}
	\end{minipage}%
	\begin{minipage}{.5\textwidth}

		\includegraphics[width=7cm]{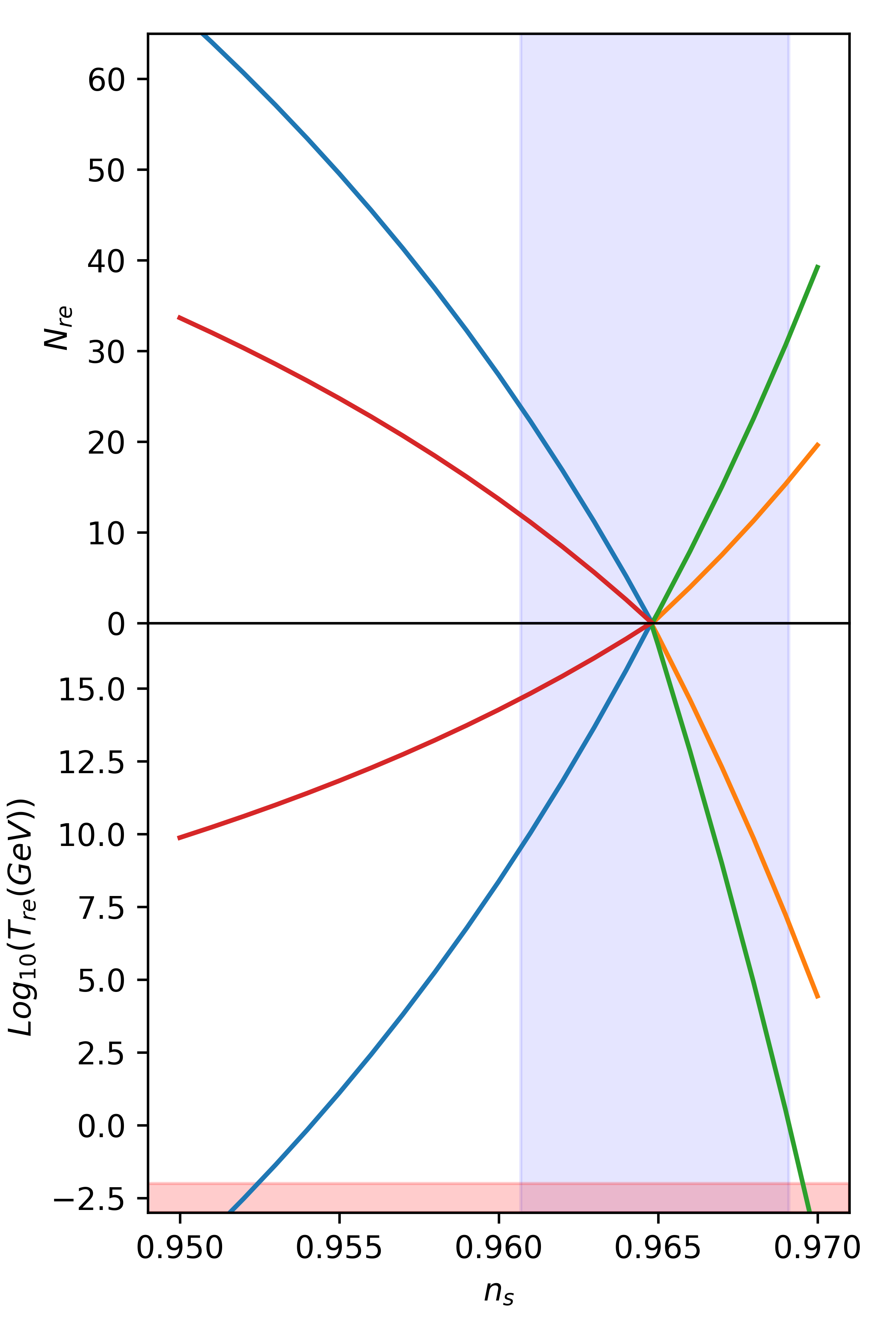}
	    
	    \captionsetup[figure]{position=bottom,justification=centering,width=.85\textwidth,labelfont=bf,font=small}
		\caption{ Plots of $N_{re}$ and $T_{re}$ with respect \\
			to $n_s$. Here red, blue, green and orange colour 
			\\
			lines correspond to $\omega_{re}=-\frac{1}{3}, 0, \frac{2}{3}$ and $1$ \\
			respectively.  }
		\label{fig:reheatingcase3change}
	\end{minipage}
	
\end{figure}

Substituting $N_{k}$, $n_s$ and $V_{end}$ into equation \eqref{nre4} and equation \eqref{tre1} and varying $n_s$ from $0.95$ to $0.97$, we plot $N_{re}$ and $T_{re}$ vs $n_s$ in the figure \ref{fig:reheatingcase3}. Here we have taken $\a =10^{-4} $, $\b=10^{-4} $ and $\c= 0.03 $. These are the same values using which we obtained figures \ref{fig:c3cvsu} and \ref{fig:c3hvsc}. It is observed that all $\omega_{re}$ values give rise to reheating parameters within known bound. If we change $\c$ by an order of $100$, we did not find any substantial change in $n_s$ value. This can be seen from figure \ref{fig:reheatingcase3change}, here $\c$ is set to $1$ and $\a$, $\b$ has same value as in figure \ref{fig:reheatingcase3}.


\subsection*{Case 4 : $V=\b h^2$ and $G=1+\c h^2$}

Similar to the case 3, here also we need to take a simplified form of $N_k$ by taking appropriate approximation. Taking hint from the values in figures \ref{fig:c4cvsu} and \ref{fig:c4hvsc}, along with the fact that value of $h_k$ is of $O(10)$, we can neglect the $\c h^2$ term in comparison to $ 8 \a \b h^4 $ and $1$ in equation \eqref{Ucae4}. With this approximation, $\epsilon$, $\eta$ and $N_k$ become -

\begin{eqnarray}
	\epsilon \simeq \frac{2}{h^2(1 + 8 \a \b h^2 )} ,
\end{eqnarray}

\begin{eqnarray}
\eta \simeq \frac{2 - 32 \a \b h^2}{h^2(1 + 8 \a \b h^2 )} ,
\end{eqnarray}

\begin{eqnarray}
	N_k \simeq \frac{h^2}{4} .
\end{eqnarray}

Using these equations, we obtain the following relations -

\begin{eqnarray}
	N_k \simeq \frac{2}{1 - n_s} ,
\end{eqnarray}  

\begin{eqnarray}
	\epsilon \simeq \frac{2}{(\frac{8}{1 - n_s})(1 + 8 \a \b \frac{8}{1- n_s})} ,
\end{eqnarray}

\begin{eqnarray}
	V_{end} \simeq 3 H^2_k \frac{(1 + 4 \c N_k)(32 \a \b N^2_k + 1 + 4 \c N_k) h^2_f}{(1 + \c h^2_f)(8 \a \b h^4_f + 1 + \c h^2_f)  4 N_K} .
\end{eqnarray}

Now we are ready to obtain $T_{re}$ and $N_{re}$ as a function of $n_s$. 

 \begin{figure}[h]
	\centering
	
	\includegraphics[width=7cm]{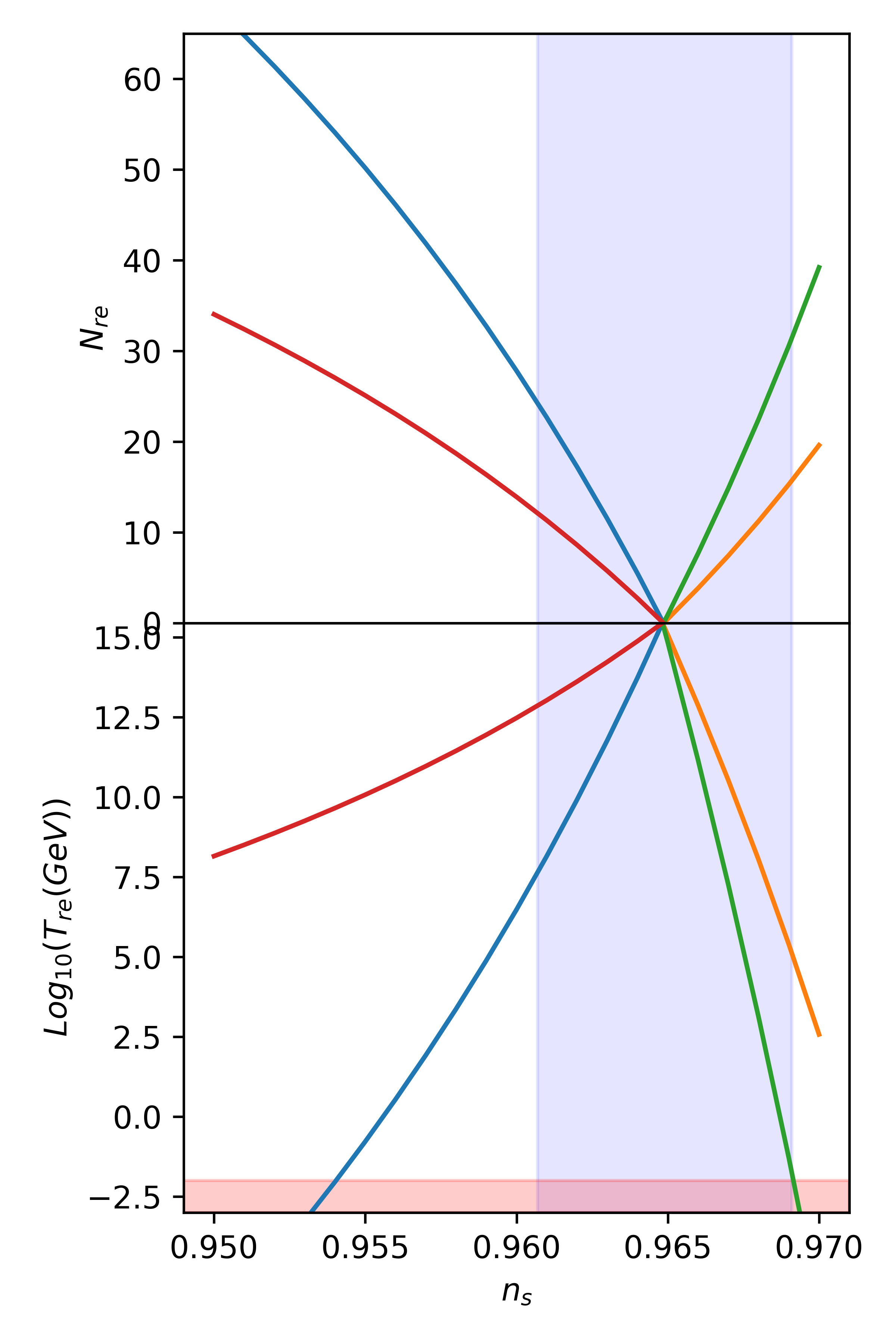}
	
	\captionsetup{justification=centering,margin=2cm}
	
	\caption{ Plots of $N_{re}$ and $T_{re}$ with respect to $n_s$. Here red, blue, green and orange colour lines correspond to $\omega_{re}=-\frac{1}{3}, 0, \frac{2}{3}$ and $1$ respectively.  }
	\label{fig:reheatingcase4}
\end{figure}

Using the same values as in figure \ref{fig:c4cvsu} (i.e. $\a=0.1$, $\b=0.01$ and $\c = 10^{-4}$ ), we obtain figure \ref{fig:reheatingcase4}. Here also, all four values of $\omega_{re}$ give rise to possibles values of $N_{re}$ and $T_{re}$.

\section{Conclusion}  \label{sec:conclusion}
Inflation models are constructed in a modified gravity theory in Palatini formalism. Here our modified action contains a general non-minimal coupling of scalar field with $R$
and $R^2$ term separately.  Unlike metric formalism, Palatini formalism does not introduce a new scalar degree of freedom in our theory. It turns out that  $R^2$ term in the action can be translated to a higher order kinetic term and a new potential term for scalar field in an equivalent scalar-tensor set-up in Einstein frame.  However, in slow-roll inflation setting we  can safely neglect the effect of higher order kinetic term in the action.  The modified potential is responsible for the inflation. In this equivalent scalar-tensor theory, we calculate scalar spectral index$(n_s)$ and tensor to scalar ratio($r$) for four different cases depending on the form of nonminimal coupling and potential of scalar field. We have varied three unknown parameters of our theory to extract the information about $n_s$ and $r.$ It is observed that in most of the cases we have a large parameter 
space which can match with the results of Planck 2018 constraints on $n_s$ and $r.$  Palatini formalism in modified gravity theories may provide a large class of models for 
Inflation satisfying CMB data. We have studied reheating phenomenon for the models considered in this paper which are compatible with Planck data. Also a more general action with nonminimal coupling of scalar field can be studied to explore the cosmology of very early universe. A dark energy picture of these models is worth investing in future.

\acknowledgments

This work was partially funded by DST (Govt. of India), Grant No. SERB/PHY/2017041.

\bibliographystyle{JHEP}
\bibliography{paper}

\end{document}